\documentclass[aps,pre,twocolumn,unsortedaddress]{revtex4}

\usepackage[version=3]{mhchem} 
\usepackage{siunitx}
\usepackage{xcolor}
\usepackage{graphicx}

\begin{document}

\author{Francis Dragulet}
\affiliation{Department of Physics, Institute for Soft Matter Synthesis and Metrology, Georgetown University, 37th and O streets NW, Washington, D.C. 20057, USA}
\author{Abhay Goyal}
\affiliation{Department of Physics, Institute for Soft Matter Synthesis and Metrology, Georgetown University, 37th and O streets NW, Washington, D.C. 20057, USA}
\affiliation{Infrastructure Materials Group, Engineering Laboratory, National Institute of Standards and Technology, 100 Bureau Drive, Gaithersburg, MD 20899}
\author{Katerina Ioannidou}
\affiliation{CNRS and University of Montpellier, 641 Av. du Doyen Gaston Giraud, 34000 Montpellier, France}
\author{Roland J.-M. Pellenq}
\affiliation{EPiDaPo, the joint CNRS and George Washington University laboratory, Children's National Medical Center, Children's Research Institute, 111 Michigan Avenue NW, Washington, D.C. 20010, USA}
\author{Emanuela Del Gado}
\affiliation{Department of Physics, Institute for Soft Matter Synthesis and Metrology, Georgetown University, 37th and O streets NW, Washington, D.C. 20057, USA}

\title{Ion specificity of confined ion-water structuring and nanoscale surface forces in clays}

\keywords{Ion specificity, ion-water confinement, self-assembly, clays, nanoscale forces, molecular dynamics, Monte Carlo}

\begin{abstract}

Ion specificity and related Hofmeister effects, ubiquitous in aqueous systems, can have spectacular consequences in hydrated clays, where ion-specific nanoscale surface forces can determine large scale cohesive, swelling and shrinkage behaviors of soil and sediments. We have used   
a semi-atomistic computational approach and examined sodium, calcium and aluminum counterions confined with water between charged surfaces representative of clay materials, to show that ion-water structuring in nanoscale confinement is at the origin of surface forces between clay particles which are intrinsically ion-specific. When charged surfaces strongly confine ions and water, the amplitude and oscillations of the net pressure naturally emerge from the interplay of electrostatics and steric effects, which can not be captured by existing theories. Increasing confinement and surface charge densities promote ion-water structures that increasingly deviate from the ions' bulk hydration shells, being strongly anisotropic and persistent, and self-organizing into optimized, nearly solid-like assemblies where hardly any free water is left. In these conditions, strongly attractive interactions can prevail between charged surfaces, due to the dramatically reduced dielectric screening of water and the highly organized water-ion structures. By unravelling the ion-specific nature of these nanoscale interactions, we provide
evidence that ion-specific solvation structures determined by confinement are at the origin of ion specificity in clays and potentially a broader range of confined aqueous systems. 
  
\end{abstract}
\maketitle
\section{Introduction}
Clay minerals are ubiquitous components in essentially all soils and sedimentary environments on Earth and other planets. They constitute a historically important, and newly rediscovered, route to sustainable and locally sourced construction materials and are key players in a wide range of geophysical phenomena including mudslides, debris flows, fault slip and ground subsidence \cite{mud-2020,deshpande2021perpetual,bourg2017clay,hochella2019natural,van2018earth,gallipoli2017geotechnical}. 
Ultimately, their sensitivity to salinity, pH, moisture and load/flow conditions originates from the nanoscale physical chemistry and ionic composition of clay layers, from which larger scale structures with complex pore networks and load bearing properties develop \cite{Seiphoori2020formation,brochard2017nanoscale}. Nanoscale surface forces in hydrated clays, therefore, play a key role in the geological disposal and management of waste, in the stability of soils and building foundations, and in their potential as a construction material, because it determines cohesion, shrinkage or swelling. This macroscopic behavior originates from the interactions between the charged surfaces of clay nanoparticles, which are intercalated with ions and water in soils with different degrees of humidity.

The governing forces between charged surfaces in solutions are described by the classical approach based on Derjaguin-Landau-Verwey-Overbeek (DLVO) theory, which uses the Poisson-Boltzmann formulation and considers the point-charged ions confined between two like-charged surfaces as a charged gas embedded in a dielectric continuum (water) \cite{israelachvili2015intermolecular,deraguin1941theory,verwey1948theory}. In many cases, this mean-field formulation captures the essential physics, however the validity of the dilute ionic gas assumption can be questioned in several cases, where discrete effects indeed change the nature of the resulting forces. Experimental measurements using surface force apparatus (SFA) and atomic force microscopy (AFM) have in fact confirmed the presence of non-DLVO forces in systems strongly confined by charged surfaces, especially for surface separations below 3 nm  \cite{israelachvili1983molecular, pashley1984molecular, pashley1981dlvo, perkin2011self, perkin2012ionic, plassard2005nanoscale,salles2010cation,zachariah2016stepwise,zachariah2017ion,espinosa2012hydrated}.
While the DLVO framework always predicts a dominating surface-to-surface electrostatic repulsion when ions are monovalent and surface charge densities are relatively small, surface forces in solutions are instead often found to be ion specific and depend non-trivially on ion valency, akin to the Hofmeister effect seen in proteins or colloids \cite{hofmeister1888understanding,kunz2004lehre,kunz2004present,Rad2015,bostrom2001specific,duignan2014collins,ninham1997ion,Innes-Gold2021,li2017strong}. The dependence of the interaction strength and sign on the ionic composition in clays, in fact, has the characteristics of a Hofmeister series, but this dependence cannot be explained within DLVO theory and is not captured by the current understanding of Hofmeister effects in clays \cite{underwood2016ion,yi2018surface}.

Non-DLVO effects have been investigated through primitive model (PM) simulations that use ions with a finite size and explicitly allow for ion-ion correlations to emerge, which reflects the fact that an excess of finite-size ions in some places in the interlayer clay void leads to a deficit in other places \cite{jonsson1980ion, kjellander1986double,kjellander1988double,jonsson2004ion,pegado2016attractive}. These ion-ion correlated density fluctuations give rise to an attractive force in the same way that correlations between instantaneous electronic dipoles (due to the quantum fluctuations of the electron density around two atoms) give rise to London dispersion forces. This non-classical feature of the double layer, inducing an attraction between two similarly charged surfaces immersed in an electrolyte, has been explored within the theory of the equilibrium electrical double layer, where two regimes can be distinguished \cite{pellenq1997electrostatic}. The first regime is an attractive regime that occurs at small surface-surface separations and in the presence of multivalent ions, where the forces change from repulsive to attractive as the surface charge and pH increase. The second regime occurs at larger separations, and especially in the case of multivalent ions, the net force can be either attractive or repulsive depending on the electrostatic coupling.

While the primitive model correctly accounts for these non-DLVO effects, the model's treatment of the water as an isotropic dielectric continuum, and its disregard for ion hydration forces, remain questionable in the regime of small separations. In strongly confined systems, in fact, the dynamics, layering, and dielectrical properties of water are known to deviate strongly from that in bulk; the dielectric constant becomes significantly reduced, strongly anisotropic, and spatially inhomogenous \cite{fumagalli2018anomalously,Shen2021,goyal2021physics,schlaich2018simulations,zhang2013strongly,Monet2021nonlocal,BenoitThesis,motevaselian2020universal,jimenez2020nonreciprocal,leoni2021nanoconfined}. 
For this reason, simulations with atomistic resolution are needed to capture important physics previously missed in clay studies, and to provide new evidence of non-DLVO forces and ion specific effects \cite{rotenberg2014multiscale,tesson2018classical,Shen2021,LeCrom2020,simonnin2018mineral,bourg2017clay,whitley2004free,young2000simulations,BenoitThesis,brochard2021swelling}. Nevertheless, due to the large parameter space to explore, and the complexity of the information to process, fully atomistic studies have yet to provide a consistent and more general picture for the emergence of ion specificity and for the origin of Hofmeister effects. To distill new fundamental understanding from the evidence provided by fully atomistic studies and nanoscale forces measurements, we turn to a semi-atomistic modeling approach that uses explicit ions and water confined between surfaces whose physical chemistry is captured by their surface charge density \cite{goyal2021physics}. This approach allows us to demonstrate that both size and ion valency determine the ion-water structures stabilized under confinement between clay surfaces, and that their interplay can explain Hofmeister effects and important changes in non-DLVO contributions to nanoscale surface forces in clays.   

Using Molecular Dynamics simulations of ions and water confined between charged planar surfaces, we examine how the cohesion and swelling properties vary with the type of charge-balancing cation. We demonstrate that in clays the source of these properties lies in the ion-water structuring that induces strong and ion-specific correlations, and that both ions and water must be accounted for to obtain a complete picture of the governing electrostatics. Specifically, we carry out this study at a surface charge density typical of smectite clays ($\sigma = 1e^{-}/nm^{2}$), and with counterions common to this family of clays (Na\(^{+}\), Ca\(^{2+}\), and Al\(^{3+}\)). We also consider a significantly higher surface charge density ($\sigma = 3e^{-}/nm^{2}$) to determine how our findings might change with surface chemistry. By using a semi-atomistic approach, we are able to efficiently sample various surface charge densities, counterion types, and a wider range of pore widths. With this approach, we can also generate numerous independent ensembles from scratch, to thoroughly and quantitatively evaluate fluctuations and statistical correlations. 

Crucially, we find that ion specificity is enhanced under strong confinement, where water structure deviates from the bulk liquid. A thorough investigation of structural and dynamical correlations between ions and water reveals that the ion-water structures and their interactions with the charged surfaces control the resulting net pressure. These findings explain how the stability and strength of hydrated clays are ion-specific. The nanoscale effects unraveled here are reminiscent of the mechanisms invoked to explain Hofmeister series in solutions of macromolecules, and in a range of biological and colloidal dispersions  \cite{zhang2006interactions,parsons2011hofmeister,kunz2010specific,andreev2018influence,ball2019hofmeister,okur2017beyond,dos2011ion,underwood2016ion,argyris2010ion}. 
Studies of Hofmeister effects in all these systems have in fact highlighted the role of ion-specific solvation and related changes close to surfaces. The ion-specific and confinement-specific solvation structures unraveled here potentially provide, therefore, the missing link between ion-solvation close to surfaces and ion-specific effects not only in clays but also, more broadly, in a range of aqueous systems.

\section{Methods}

In order to examine ion-specific effects in clays, we performed simulations with three ion types (Na\(^{+}\), Ca\(^{2+}\), and Al\(^{3+}\)) over a range of surface charges from $\sigma = 1e^{-}/nm^{2}$ to $\sigma = 3e^{-}/nm^{2}$ and surface separations from $D=6$\si{\angstrom} to $D=40$\si{\angstrom}. We sample this large parameter space with a semi-atomistic approach that captures the important effects missed by DLVO theory or the primitive model (PM). For each data point, all quantities have been averaged over $10$ statistically independent samples and, for each of these samples, over $10^{6}$ MD steps after reaching equilibrium (more details below). The error bars obtained from the sample-to-sample fluctuations are less than, or equal to, the size of the symbols in the figures.

\subsubsection*{Semi-atomistic approach}
In our computational model, the ions are confined to a slab which is finite in the $\hat{z}$ direction and periodic in $\hat{x}$ and $\hat{y}$ (the directions parallel to the surfaces). The clay surfaces are treated as infinite, uniformly-charged walls with a characteristic surface charge density $\sigma$, varying from $1 e^{-}/nm^{2}$ to $3 e^{-}/nm^{2}$. This description of the confining surfaces allows us to perform simulations that can extensively sample the microstates of the confined ions and water, and therefore to extract the spatio-temporal correlations required to determine the microscopic origin of the nanoscale forces. Previous work has demonstrated that the physical mechanisms dictating the cohesion of charged layers in an ionic solvent can be fundamentally captured by a non-atomistic treatment of the walls, at a much lower computational cost \cite{BenoitThesis,goyal2021physics}.

To keep all systems examined charge neutral, the number of counterions, \(N_{ion}\), and the area of the plates, \(L_{x}L_{y}\), are adjusted (see table \ref{simParam}).
\begin{table*}
  \caption{Simulation Parameters}
  \label{simParam}
  \begin{tabular}{cccccc}
    \hline
    Ion/Atom & $\sigma$ (\(e^{-}/nm^{2}\)) & \(N_{ion}\) & \(L_{x}\)=\(L_{y}\)(\si{\angstrom}) & $\epsilon$ (kcal/mol) & d(\si{\angstrom})  \\
    \hline
    Na\(^{+}\) & 1 & 128 & 80 & 0.1301\textsuperscript{\emph{a}} & 2.350\textsuperscript{\emph{a}}    \\
     & 3 &  & 46.188  \\
    Ca\(^{2+}\) & 1 & 64 & 80 & 0.1000\textsuperscript{\emph{a}} & 2.870\textsuperscript{\emph{a}}  \\
    & 3 &  & 46.188  \\
    Al\(^{3+}\) & 1 & 42 & 79.373 & 0.2166\textsuperscript{\emph{b}} & 1.447\textsuperscript{\emph{b}}  \\
    & 3 &  & 45.826  \\
    O &  &  &  & 0.1554\textsuperscript{\emph{a}} & 3.166\textsuperscript{\emph{a}}    \\
    \hline
  \end{tabular}
  
  \textsuperscript{\emph{a}} \citeauthor{cygan2004molecular};
  \textsuperscript{\emph{b}} \citeauthor{faro2010lennard}
\end{table*}
The dispersion and steric interactions between particle \(i\) and particle \(j\) are described by a 12-6 Lennard-Jones potential: 

\begin{equation}
U^{LJ}(r_{ij})=4\epsilon_{ij} \left[ \left(\frac{d_{ij}}{r_{ij}} \right)^{12}- \left(\frac{d_{ij}}{r_{ij}} \right)^{6} \right] 
\end{equation}
which is cutoff and shifted at a distance of 20\si{\angstrom}. For interactions involving water molecules, the much larger oxygen atom in the SPC/E water model serves as the LJ site. We use the SPC/E model for the water as it accurately matches experimental data on the density, structure and dynamics of bulk water at room temperature \cite{berendsen1987missing,mark2001structure}. A recent study indicated that using another rigid water model, such as TIP4P, did not significantly affect the cohesion or other behavior in confinement \cite{goyal2021physics}. Further, clay studies with polarizable water models have shown that the inclusion of polarizability results in less dynamic water \cite{LeCrom2020}, while examining confined water between MgO surfaces reveals longer-ranged water layering with polarizable water \cite{kamath2013comparison}. This suggests that polarizable water would work to accentuate our results, and their greater computational cost serves to justify our choice of the SPC/E water model.

The parameters for the Lennard-Jones potential are taken from \citeauthor{cygan2004molecular} and \citeauthor{faro2010lennard}, and shown in table \ref{simParam}. Cross-species interactions are described by the arithmetic average of the two parameters (i.e. $d_{ij}$ = ($d_{i}+d_{j}$)/2). The interactions with the walls are quantified in a similar fashion and all ions and water molecules that interact with the fixed boundaries in the z-direction experience a LJ force perpendicular to the wall. The LJ parameters for the walls are the same as that used for water oxygen; using a different value for \(d\) would simply translate to shifting the effective plate separation, or pore size. 

Coulomb forces are used to describe the electrostatic interactions between the atoms in our simulation. Here, the charge of the ions is simply given by its valency (i.e. Al\(^{3+}\) has a charge of +3e). To account for the long-ranged nature of Coulombic forces, the summation of the electrostatic forces across periodic images is reliably and efficiently computed by the Ewald method \cite{allen2017computer}. For our 2D periodic slab geometry, the classic Ewald summation is modified by including a correction term: \cite{yeh1999ewald} 

\begin{equation}
E_{slab}(M) = \frac{2\pi}{V}M_{z}^{2}
\end{equation}
where \(M_{z}\) is the z component of the total dipole moment of the cell \cite{smith1981electrostatic}.

With this semi-atomistic approach, Molecular Dynamics simulations are performed in the NVT ensemble, with a timestep of 1 fs and with the system kept at room temperature via a Nos\-e-Hoover thermostat. All simulations are done using LAMMPS \cite{plimpton1995fast}, and are performed for surface-surface separations ranging from D=6\si{\angstrom} to D=40\si{\angstrom}. In the high surface charge density case ($\sigma=3 e^{-}/nm^{2}$), simulations are also conducted at even smaller separations, to ensure that the pressure does not strictly become more attractive as D decreases. Additionally, multiple independent runs are carried out at the same value of D to verify that our results are not unique to a particular ensemble. Below, we provide further details on how these simulations and other calculations are performed.   

\subsubsection*{Water insertion and reaching equilibrium}

The first step in our computational procedure is to assign each ion a random starting position within the simulation cell. With this initial configuration, water molecules are inserted into the simulation box via Grand Canonical Monte Carlo (GCMC) simulations, where a water molecule is inserted at a randomly selected position with a probability determined by the Metropolis method \cite{frenkel2001understanding}. The value of the chemical potential used in the Metropolis method, \(\mu\) = -8.8 kcal/mol, was determined by conducting simulations in bulk water conditions \cite{goyal2021physics}. The acceptance rate of inserting water molecules decreases as more water molecules are added, which significantly slows down the process of approaching equilibrium, or a constant density of water. This process was sped up by conducting Molecular Dynamics while this GCMC process was taking place \cite{BenoitThesis}. By allowing the ions and newly introduced water molecules to move, more space is ``freed up," thereby increasing the probability of inserting a new water molecule. The time required to reach equilibrium is longer at lower surface charge densities and larger simulation boxes. The convergence was improved by inserting water molecules to match the density of bulk water (1 g/cm\(^{3}\)) prior to starting the GCMC process, as this is closer to the desired final density. Regardless of the initial configuration, $1000$ particle exchanges (insertions or deletions) are attempted every $1000$ MD timesteps until the water density remains constant (up to 6*10\(^{6}\) MD steps). The final number of water molecules grows with the system size, and goes up to approximately 24,000 (for $\sigma=1 e^{-}/nm^{2}$ and D=40\si{\angstrom}), which is an order of magnitude larger than the number of water molecules used in recent fully-atomistic simulations \cite{simonnin2018mineral,LeCrom2020,brochard2021swelling}.

With the final configuration generated by this combined GCMC/MD water insertion process, an equilibration Molecular Dynamics run was conducted. Equilibrium was determined by verifying that the pressure fluctuates around a mean value, and that our time-correlation functions (MSD, bond correlation function, scattering function) do not depend on the choice of the initial timestep, t\(_{0}\). Our confidence in reaching equilibrium was also raised by noting that the results do not change with examining numerous independent simulations at the same surface-surface separation. The duration of our equilibration runs is typically 10\(^{5}\) timesteps. Overall, this entire methodology also applies to the primitive model (PM) simulations, but with no water molecules inserted and the dielectric constant set to the approximate bulk value for water at room temperature ($\epsilon_{bulk}$ = 78). 

\subsubsection*{Observables in equilibrium}

After equilibrium is reached, a production run of 10\(^{6}\) timesteps follows. For each independent run at each value of D, the pressure between the two walls, or the z-component of the stress tensor, is time-averaged over this production run as the force exerted on the charged surfaces. The pressure at a bulk-like separation (D=40\si{\angstrom}) was subtracted from the pressure at each value of D to produce the curves in the figures. The final pressure, as well as the quantities that follow, at each value of D were computed as the average over 10 independently-created ensembles, or simulations. For these independent simulations, water insertion runs, equilibration runs, and production runs were all carried out from a different starting configuration of ions. At separations of D$\leq$12\si{\angstrom}, the errors in pressure, coordination number, and dielectric constant are smaller than the symbols used in the figures. The same holds for the density profiles, $g(r)$, angle distributions, and correlation functions shown in the paper. For larger separations, D$\geq$13\si{\angstrom}, we use only time averages, since dynamics is faster and overall time correlation functions decay over the simulations time window (see also discussion below). 

To gain further insight into the origins of the pressure and its variations, we examined how the explicit water's ability to screen charges depends on the type of counterion. This is measured by the dielectric constant, which can be computed in molecular simulations from the total dipole moment, M, via the fluctuation-dissipation theorem \cite{gray1986computer}. As the system is anisotropic, the dielectric properties are described by a tensor rather than a single value. By symmetry, the $x$ and $y$ components are equal, and $\epsilon_{xy}$ is related to the x and y components of M by:\cite{froltsov2007dielectric}

\begin{equation}
    \epsilon_{xy}=1+\frac{1}{\epsilon_{0}Vk_{b}T} \frac{\langle M_{x}^{2} + M_{y}^{2} \rangle}{2}
\end{equation}

We do not compute $\epsilon_z$ because the slow dynamics in the $z$ direction require longer simulation times than we ran in order to compute it accurately. The structure of the ions, and their positional correlations, is examined by computing the pair correlation function, or \(g(r)\) in the plane parallel to the surfaces. It is defined as 
\begin{equation}
    g(r)=\frac{L_{x}L_{y}}{2 \pi r \Delta r \: N^{2}} \langle \sum_{i}^{N} \sum_{j \neq i}^{N} \Theta(\Delta r - |r-r_{ij}|)\rangle
\end{equation}
where $\Delta r$ is size of the sampling bin, $r_{ij}$ is the distance between ions $i$ and $j$, and $\Theta$ is the Heaviside step function. This quantity is calculated for ions depending on their $z$ position, or the layer in which they reside, with N being adjusted accordingly. As with the dielectric constant, the \(g(r)\) was averaged over the whole production run ($10^{6}$ MD steps).

The strength of ion-water interactions, and thereby the stability of the hydration shells, was evaluated by measuring the time correlations of the ion-water bonds, or \(c_{iw}(t)\):

\begin{equation}
    c_{iw}(t')= \frac{\langle \Theta_{t=t'}(r_{cut}-\Delta r_{iw}) \: \Theta_{t=0}(r_{cut}-\Delta r_{iw}) \rangle}{\langle \Theta_{t=0}(r_{cut}-\Delta r_{iw})\rangle}
\end{equation}
Here, $\Theta$ is the Heaviside step function, $r_{cut}$ is the distance from the ion's center that encapsulates the bound water molecules, and $\Delta r_{iw}$ is the ion-water distance. This quantity is averaged over ions and over bound water molecules. For the sodium-water bonds, which die off by the end of the simulation, the mean relaxation time of these bonds can be estimated as: 

\begin{equation}
    \langle \tau \rangle \simeq \int_{t=0}^{t=10^{6} fs} c_{iw}(t) dt
\end{equation}

To quantify the changes in shape of the aluminum 6-mers, we calculate the bound water dipole angle distribution:
\begin{equation}
    f(\theta)= \frac{1}{N_{bound}}\langle \sum_{\theta=\ang{0}}^{\theta=\ang{180}} \frac{N_{bound}(\theta=\theta')}{sin(\theta)}\rangle
\end{equation}
with
\begin{equation}
    \theta'= \cos^{-1}{(\pm \frac{\mu_{z}}{\mu})}
\end{equation}
This entails sampling over the bound water molecules (\(N_{bound}\)) and computing the angle (\(\theta'\)), rounded to the nearest integer, between the dipole moments of the bound water (\(\mu\)) and the normal of the nearest surface (\(\mu_{z}\)). Visually, this is the angle between the yellow arrows and the z-axis, shown in Figure \ref{figure3}. This was again averaged over the production run. 

Finally, several functions are computed to detail the dynamics of the ions and water molecules, most of which is displayed in the supporting figures. The calculated quantities include the mean-square displacement in the xy-plane (MSD\(_{xy}\)), the mean-square displacement along the direction perpendicular to the surfaces (MSD\(_{z}\)), and the self-intermediate scattering function \cite{mcdonald2006theory} in the z-direction, \(F_{s}(q,t)\). These three quantities are computed as:

\begin{equation}
    \text{MSD\(_{xy}\)(t)} = \frac{1}{N} \sum_{j=1}^{N} [(y_{j}(t)-y_{j}(0))^{2}+(x_{j}(t)-x_{j}(0))^{2}]
\end{equation}

\begin{equation}
    \text{MSD\(_{z}\)(t)} = \frac{1}{N} \sum_{j=1}^{N} [(z_{j}(t)-z_{j}(0))^{2}]
\end{equation}

\begin{equation}
    F_{s}(q,t)=\frac{1}{N} \sum_{j=1}^{N} e^{i q_{z} (z_{j}(t)-z_{j}(0))}
\end{equation}
where the scattering vector component \(q_{z}\) ranges from its smallest possible value, \(q_{z} = 2 \pi / D\) to \(q_{z} = 10 \si{\angstrom}^{-1}\). All of the above time-dependent observables are averaged over all ions/water molecules. The mean-square displacement was also averaged over 10 different starting times (t\(_{0}\)) spaced out across the production run. For all these time correlation functions, we have measured their dependence on the initial time at which the measurement starts during the simulations and verified that the data do not show signs of aging. At the larger separations, even in the case of aluminum ions for which the bonds with water are strongly persistent, the ion mean squared displacements and the intermediate scattering functions indicate that equilibrium states are efficiently sampled by performing time averages.

\section{Results and discussion}

\subsection*{Effect of Explicit Water}

\begin{figure*}
\centering
\includegraphics[width=6.5in]{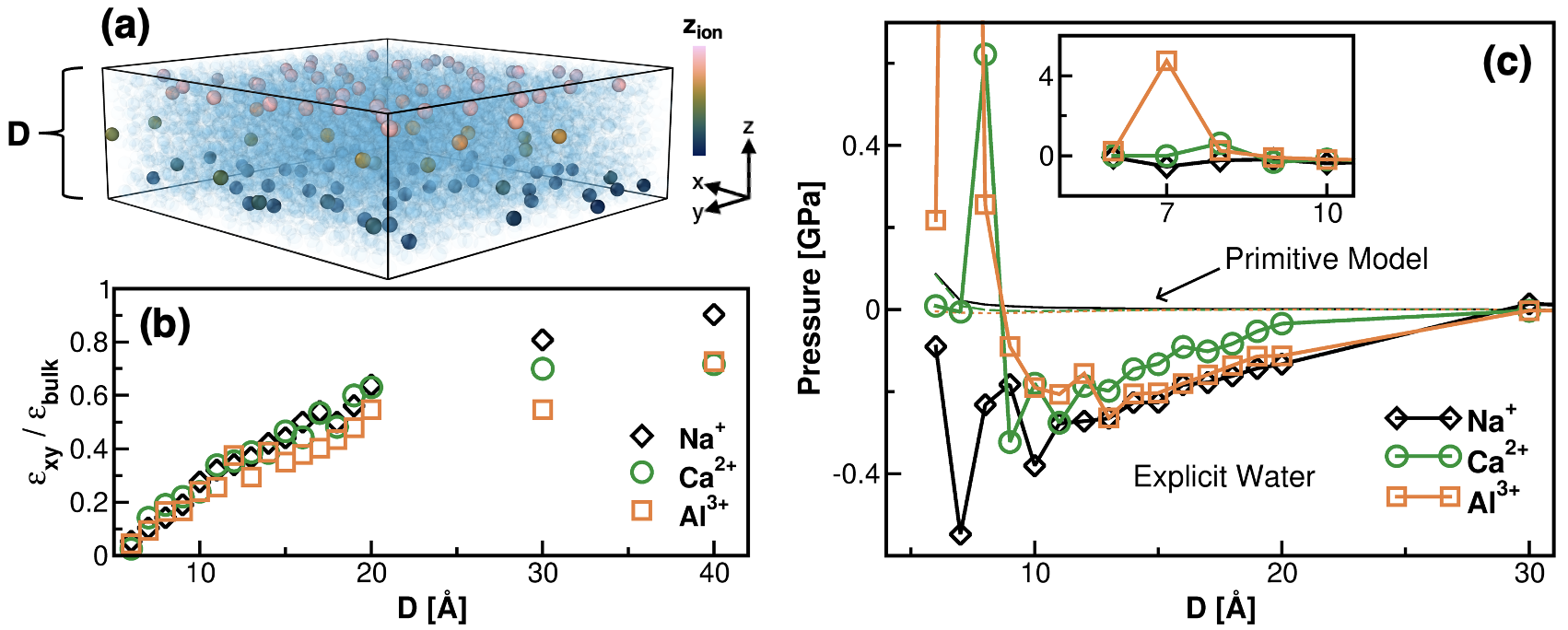}
\caption{A visualization of the system is provided in (a), where Na\(^{+}\) ions are colored by z-position and the water molecules are represented as transparent blue. For small pores, where most of the water is bound to the ions and thus unable to act effectively as a screening solvent, the dielectric constant in the x and y directions differs significantly from its z component, and is notably less than the isotropic dielectric of bulk water used in the primitive model (PM), as shown in (b). While the dielectric's dependence on pore size is similar for all three ions, the resulting net pressure between the charged surfaces (\(\sigma = 1e^{-}/nm^{2}\)) for the explicit water approach (EW), in (c), changes with the valency, ion size, and pore size in a more complex fashion. This demonstrates that utilizing EW does more than just scale down the electrostatic screening. Inset in (c): zoom showing peak in Al\(^{3+}\) pressure. }
\label{figure1}
\end{figure*}

To understand the effect of using explicit water (EW), we start by comparing our results with EW to PM simulations for $\sigma = 1e^{-}/nm^{2}$. Utilizing EW leads to ion-water binding, which limits the ability for the water dipole moment to reorient. This effect results in reduced dielectric screening ability when a large portion of the water is bound, as previously observed in experiments and simulations \cite{schlaich2018simulations, fumagalli2018anomalously,Monet2021nonlocal,goyal2021physics}. We observe a similar trend whereby increased confinement leads to lowered dielectric properties (see Figure \ref{figure1}b). In small pores, where most of the water is bound, the computed dielectric constant in the plane parallel to the surfaces (\(\epsilon_{xy}\)) is significantly less than the bulk dielectric (\(\epsilon_{bulk}\)), which unlike our calculated quantity, is isotropic. As the pore size is increased, more free water is introduced, and the less constrained water is more akin to the dielectric continuum representation utilized in the primitive model. 

While the relationship between the dielectric properties and pore size does not notably change from ion to ion, the net interaction pressure in EW varies with the ion type, and the separation, in a nontrivial way (see Figure \ref{figure1}c). This is especially true at higher confinement (D $\leq$ 12\si{\angstrom}). We note that all the data in this regime are averaged over ten independently-generated samples and that the statistical error estimated from the sample-to-sample fluctuations is smaller than the symbol sizes in the plots. In the regime of high confinement, the molecular description of water in EW induces an oscillating pressure that can be attractive at some pore widths, which are consistent with a number of previous studies of nanoscale cohesion in hydrated clays \cite{Shen2021,whitley2004free,BenoitThesis,brochard2021swelling}. With implicit water (PM), the pressure is relatively weak, does not exhibit large fluctuations with $D$, and strictly decreases with an increase in the valency of the ions. Neither of these characteristics apply to EW results. These clear qualitative differences in the pressure curves (between PM and EW as well as between ion types) make it clear that utilizing discrete water molecules has a more profound effect than a simple rescaling of the dielectric constant which could be included in a suitably modified PM model approach. Such a rescaling will only lead to a pressure curve with one minimum \cite{palaia2019charged}. Additionally, the first two minimums in the EW pressure curves for Na$^{+}$ and Ca$^{2+}$ match the experimentally observed stable states in the smectite structure, at d-spacings of about 12.5\si{\angstrom} (one-layer hydrate) and 15\si{\angstrom} (two-layer hydrate) \cite{hensen2002clays,ferrage2016investigation}, when taking into account the clay layer thickness ($\approx$ 6\si{\angstrom}).

\begin{figure*}
\centering
\includegraphics[width=6.5in]{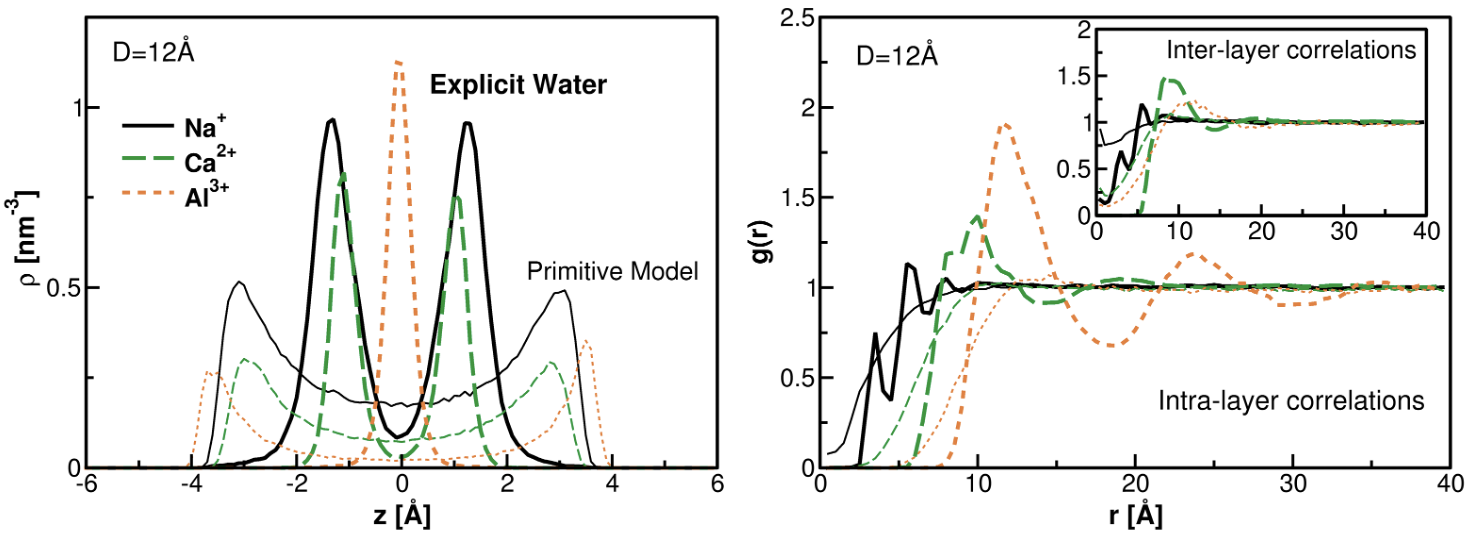}
\caption{Shown left is the densify profile of the counterions along the z-axis, at a separation of 12\si{\angstrom} and surface charge density \(1e^{-}/nm^{2}\). With implicit water (thin lines), the ions are arranged into two layers near the surfaces. When the screening is reduced with the inclusion of explicit water (thick lines), the ions get closer together; the two planes of localization for sodium and calcium are near the center while aluminum ions are united in a single central layer. The xy pair correlation function $g(r)$ on the right measures the correlations between ions along the x and y directions within these layers, and across opposite layers (inset). Positional correlations are dramatically stronger with EW, and increase with the charge of the ions. Counterintuitively however, the stronger correlations at higher valency do not coincide with an increase in the net attraction (Figure \ref{figure1}c). }
\label{figure2}
\end{figure*}

To investigate the origin of the changes in the pressure, we first examine the structuring of the ions, since a key finding of the PM approach (relative to DLVO) was that ion-ion correlations substantially affect the pressure. The ion density distributions along the z-direction (the direction perpendicular to the surfaces), quantified by ion density profiles for both the PM and EW models, are shown in Figure \ref{figure2}a. At larger separations, all ions are arranged into two clearly distinct layers, parallel to and near the charged surfaces, but the situation is more complex when the surfaces are brought closer together. At $D=12\si{\angstrom}$, for example, the ions are primarily located in two layers, with the smaller ions able to come closer to the surfaces. However, the ions are free to move across these two layers, as suggested by the density profile height between the two peaks, and the ions mean squared displacement (see Figure S1). The ions become more immobile and correlated in position as the valency is increased. This is reflected in the radial distribution functions $g(r)$ of Figure \ref{figure2}b, which measure intra-layer and inter-layer correlations between ions in the x-y plane. The inter-layer correlations, plotted in the inset, are indeed weakest for sodium and strongest for aluminum. Overall, in the implicit water framework, the ions are disordered within and across layers, but the correlations do increase slightly with the valency.

With EW, sodium and calcium ions are also arranged in two layers at D=12\si{\angstrom}. However, the ions within these layers are more strongly localized, and the two layers are closer to the midplane than their PM counterparts. Additionally, the inter- and intra-layer correlations still increase with the valency but are dramatically stronger with EW, hinting that the structuring and correlations of the ions depends on their interactions with the water. Another notable feature is the localization of Al\(^{3+}\) ions into a single central layer. At first, this appears to be in contradiction with the fact that aluminum has the highest valency: one might expect stronger electrostatic repulsion between the ions to drive the ions further away from one another. In fact, we discover that this happens precisely because of the high valency when we examine the ion-water structuring in detail in the next section.

Before getting into that, we note that overall the stronger correlations suggest the presence of stronger ion correlation forces at higher valency, which are expected to increase the attraction between the charged surfaces \cite{israelachvili2015intermolecular, jonsson1980ion, kjellander1986double}. The pressure data (see Figure \ref{figure1}c) show that while true in the PM, this is not the full picture in the EW model. To further examine the nature of the pressure, and to clarify the differences seen in the structure, we must analyze how the structuring of  water molecules around the ions.

\subsection*{Ion-water Structuring}

\begin{figure*}
\centering
\includegraphics[width=6.5in]{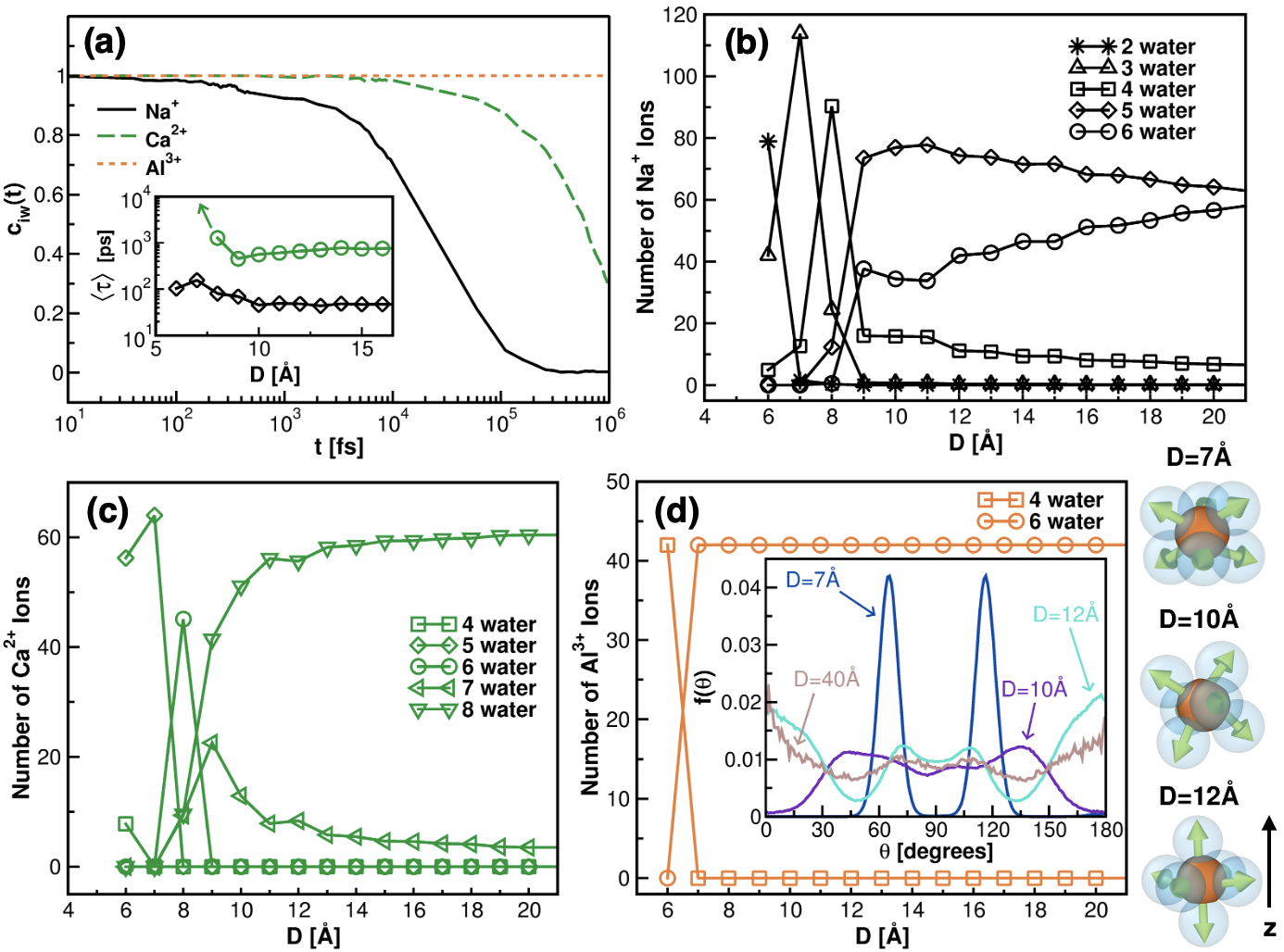}
\caption{The time correlation of ion-water bonds at D=20\si{\angstrom}, and the relaxation time for Na\(^{+}\)-water and Ca\(^{2+}\)-water bonds at various D, is plotted in (a). Water initially bound to Al\(^{3+}\) ions remains bound indefinitely, while the relaxation time for Na\(^{+}\)-water bonds is generally an order of magnitude less than Ca\(^{2+}\)-water bonds, implying that sodium hydration shells are the least persistent. However, where the Na\(^{+}\) pressure is at its minimum, D=7\si{\angstrom}, the sodium \emph{n}-mers are considerably more enduring. Plots (b) and (c) reveal sodium and calcium have full hydration shells of 5-6 and 7-8 water molecules, respectively, at large separations. Increasing the confinement, there are key values of the pore size where the size of the shells notably changes, corresponding to dips and spikes in the pressure. Aluminum hydration shells, being the most persistent, are able to endure at lower separations by exerting a stronger push against the confining wells; the persistence of these \emph{n}-mers is why the pressure becomes less attractive at higher valency for \(\sigma = 1e^{-}/nm^{2}\). Even so, the distribution of the bond angles, or shape of the Al\(^{3+}\) 6-mers, does alter significantly (see (d)) and is also correlated with fluctuations in the pressure. }
\label{figure3}
\end{figure*}

In Figure \ref{figure3}a, we plot the time correlation of the ion-water bonds, measured from our simulation data (see Methods), at a surface separation of 20\si{\angstrom} and surface charge density of \(\sigma = 1e^{-}/nm^{2}\). As the valency is increased, or as the ion-dipole attraction is stronger, water initially bound to an ion remains statistically bound for a longer period of time; Al\(^{3+}\)-water bonds tend to persist strongly well beyond the duration of the simulation. The time correlations of the bonds between the ions and the water molecules depends on time following a stretched exponential decay (Figure \ref{figure3}a). For sodium, for which the correlations decay to zero within the simulation time window for all separations $D$, the relaxation time of the Na\(^{+}\)-water bonds can be extracted as the integral of the curve. For the bond-correlation functions for the calcium-water case, we extract the relaxation time using a fit of the data available to extrapolate the relaxation to longer times, as the correlations do not fully decay within the timescales explored, after verifying that no significant aging was present (see Methods). The relaxation times for the water-ion bonds in these two cases are plotted in the inset of Figure \ref{figure3}a. 

The magnitude of bond relaxation times at larger separations is consistent with past simulations of Na\(^{+}\) and Ca\(^{2+}\) ions in bulk water \cite{israelachvili2015intermolecular,koneshan1998solvent}. When the surface separation becomes smaller than 10\si{\angstrom}, we note that the relaxation time for Na\(^{+}\) increases and reaches a maximum at D=7\si{\angstrom}. The differences in the ion-dipole attraction also manifest in the dynamical differences of the two populations of water in the parallel and perpendicular directions. The difference in mobility between the bound and free water increases with the ion charge, or ion hydration enthalpy (see figures S2-S4). This phenomena is also observed in quasielastic and inelastic neutron scattering studies of water dynamics in smectite clays \cite{swenson2000quasielastic,michot2012anisotropic,marry2013anisotropy,cygan2015inelastic}.

The number of water molecules bound to each ion depends on the size of the ions and the space available for hydration, and we describe these hydration shells as \emph{n}-mers (\emph{n} being the number of water molecules bound to an ion). Figures \ref{figure3}b, c, and d illustrate how the size of the hydration shells changes with the surface separation for sodium, calcium and aluminum, respectively. At larger separations, the size of the \emph{n}-mers is 5 or 6 water molecules for Na\(^{+}\), 8 for Ca\(^{2+}\), and 6 for Al\(^{3+}\), which is also in agreement with past bulk water simulations and experiments \cite{israelachvili2015intermolecular,galib2017revisiting, koneshan1998solvent, faro2010lennard}. The larger Ca\(^{2+}\) hydration shells and similarly-sized Na\(^{+}\) and Al\(^{3+}\) shells at these separations can explain the pressure differences at separations between 14\si{\angstrom} and 20\si{\angstrom}: larger and more stable \emph{n}-mers can enhance the repulsive nature of the hydration force, especially with increasing confinement.

As the surfaces are brought closer together, the repulsion between the ions increases, resulting in stronger ion-ion correlations. However, bringing the surfaces closer to one another also tests the stability of the full hydration shells; at key values of the separation, such as D=8\si{\angstrom} for sodium, the confining walls force the hydration shells for sodium and calcium to change in size, changes which coincide with an oscillation in the pressure (see Figure \ref{figure1}c). The clearest example of this is the reduction of \emph{n}=6 for Al\(^{3+}\) at D=7\si{\angstrom} to \emph{n}=4 at D=6\si{\angstrom}. The implications of this transition for the net pressure can be roughly estimated by using the value of the Al\(^{3+}\) hydration enthalpy, \(\simeq\)-4700 kJ/mol \cite{koneshan1998solvent}. This value indicates that an energetic cost of \(\simeq\)640\(k_{b}T\) is needed to reduce the size of one Al\(^{3+}\) hydration shell by 2 water molecules, translating to a total pressure increase of \(\simeq\)3 GPa. While this calculation does not provide the correct value of the pressure change (see peak in inset of Figure \ref{figure1}c), as it ignores the water-water and water-wall steric interactions, as well as the presence of a second hydration shell for Al\(^{3+}\) at D=7\si{\angstrom}, it nonetheless illustrates the large amount of energy needed to break up the \emph{n}-mers. Another issue in this case is that the strong Coulombic forces in reality can drive water molecules to ionize into a free H\(^{+}\) and an OH\(^{-}\) group attached to the Al\(^{3+}\), an effect that cannot be captured by our model. Nonetheless, as discussed in the literature \cite{legg2020visualization}, this ionization does not significantly change the ion-water (or ion-OH) coordination, which matches the confinement-dependent 6-mers or 4-mers that we observe.

\begin{figure*}
\centering
\includegraphics[width=6.5in]{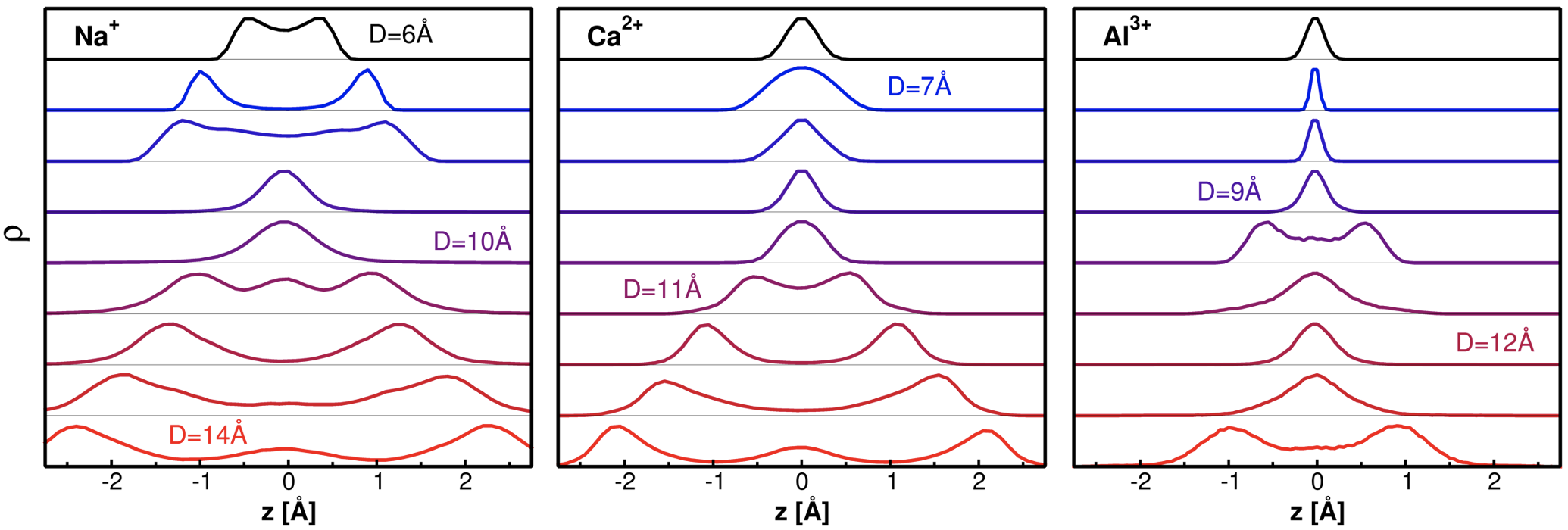}
\caption{The density profiles of sodium (left), calcium (center), and aluminum (right) ions from D=14\si{\angstrom} (red) to D=7\si{\angstrom} (black) reveal the non-monotonic relationship between pore size and layering (all profiles here are scaled to have the same height for visual clarity). At large separations, ions generally prefer to be in two layers due to their mutual repulsion, but they can be forced into a single layer due to ion-water attractions or the stability and size of the \emph{n}-mers. For sodium and aluminum ions, these competing effects lead to both coalescence and re-splitting of the ion density peaks as confinement is increased. Calcium ions, which are the largest in size, do not exhibit this behavior and have a single change from two layers at D=11\si{\angstrom} to one at D=10\si{\angstrom}. Overall, this phenomenon is not seen with the primitive model, and it helps clarify observed variations in the pressure (see text).}
\label{figure4}
\end{figure*}

Increasing the confinement while testing the stability of the \emph{n}-mers also alters the layering of the ions, and thereby, the ion-surface interactions. Notably, the variations in ion structures along the z-direction depend non-monotonically on the surface separation for sodium and aluminum, as shown by the rescaled and shifted ion density profiles in Figure \ref{figure4}. Additionally, we see that the profiles for Na\(^{+}\) ions are more spread out than that of Al\(^{3+}\). This signals that the Na\(^{+}\) ions are able to get closer to the surface, which is consistent with previous results suggesting that weakly-hydrated ions are more likely to be situated closer to clay surfaces then strongly-hydrated ions \cite{underwood2016ion}.  The relationship between \emph{n}-mer stability and these density profiles provide insight into the behavior of the pressure, and we will examine these dependencies by first looking at their interplay for Na\(^{+}\) ions.

Figures \ref{figure3}b and \ref{figure4} indicate that sodium ions have bulk water-like hydration shells and are positioned in two layers until the separation is lowered to about 10\si{\angstrom}. Here, there is no longer room for the bulk-like 6-mers and 5-mers to reside in two layers. At D=9\si{\angstrom}, the 6-mers and 5-mers are squeezed into just one layer, exerting a stronger push against the walls to maintain their energetically-favorable shell, and resulting in an uptick in the net pressure as seen in Figure \ref{figure1}c. Increasing the confinement further, at D=8\si{\angstrom} most Na\(^{+}\) ions are now bound to 4 water molecules and are spread out across the z-direction. For a separation of 7\si{\angstrom}, the system is too confined for a 4th bound water molecule and the ions localize into two distinct layers pressed onto the surfaces. The water in these 3-mers have the longest bond relaxation time (Figure \ref{figure3}a), and since a large majority of the water at D=7\si{\angstrom} is bound, the water here provides minimal dielectric screening. Because of this, and because the 3-mers exert a minimal push against the surfaces for extra space, a minimum in the pressure is reached. The pressure rises again at D=6\si{\angstrom}, as the effective pore size approaches the diameter of the ions.  

A similar chain of reasoning can be applied to the other two ion species. Divalent calcium ions, because of the larger size and stronger electrostatic interactions with water than the sodium ones, have larger and more stable hydration shells. Therefore, after the ions are squeezed into a single layer (at D=11\si{\angstrom}), there is not enough room at greater confinement for the ions to split  into two layers with smaller hydration shells, as it was the case with sodium (see Figure \ref{figure4}). Nevertheless, reducing the pore size from D=9\si{\angstrom} to 7\si{\angstrom} requires overcoming the repulsive barrier associated with reducing \emph{n} from 8 to 5 water molecules, causing a spike in the pressure at D=8\si{\angstrom}. Also noteworthy is the bond relaxation time at low D (Figure \ref{figure3}a); the lifetime of the Ca\(^{2+}\) \emph{n}-mers is the lowest at D=9\si{\angstrom}. The pressure is also at a minimum at this separation, suggesting that the reduced persistence of the \emph{n}-mers bonds at D=9\si{\angstrom} may reduce the repulsive hydration contribution to the net surface-surface interaction. At D=6\si{\angstrom} and 7\si{\angstrom}, the bond correlation function c\(_{iw}(t)\) does not decay enough to extract a finite bond lifetime from our simulations. 

Of all three counterions, while having the smallest size, Al\(^{3+}\) has the highest charge, allowing the stable 6-mers found in bulk water to persist down to the smallest separation (6\si{\angstrom}) considered here. Interestingly, oscillations in the pressure are still present even though the water coordination of the ions does not change. Unlike Na\(^{+}\) and Ca\(^{2+}\), peaks and troughs in the pressure for Al\(^{3+}\) coincide with changes in the shape of the 6-mers, and its relation to ion layering. To quantify the variations in morphology, in the inset of Figure \ref{figure3}d we have plotted the distribution of the angle $\theta$ that the dipole moment of the bound water molecules forms with the $\hat{z}$ direction for different values of D, and we also provide a visualizations of the corresponding $n$-mers shapes. When the single layer of Al\(^{3+}\) ions is confined from D=13\si{\angstrom} to D=12\si{\angstrom}, the net pressure between the surfaces increases, as the 6-mers stretch along the $\hat{z}$ direction to preserve their configuration, evidenced by a maximum in the number of water dipole moments oriented towards the surfaces (angle $\theta$ values of 0 and 180 degrees). Analogous to the case of sodium, a local minimum in the pressure (around D$\simeq 10-11\si{\angstrom}$) coincides with the ions returning to a single layer or to two layers, which disfavor the 6-mers preferred orientation of having the water dipoles point towards the surfaces ($\theta$ values of 0 and 180 degrees are drastically reduced). Upon further confinement, the ions go back into a single layer, and, at D=8\si{\angstrom} and 7\si{\angstrom}, the energetically favorable 6-mers are forced into a compact and rigid arrangement with two preferred dipole moment angles tilted with respect to the surfaces. Reducing the plate separation from D=8\si{\angstrom} to D=6\si{\angstrom} necessitates overcoming the aforementioned repulsive barrier due to the 6-mers strong tendency to preserve their state. 

To summarize, at a low surface charge density, the stability of the \emph{n}-mers appears to be a crucial determinant of the pressure at higher confinement. The stability of the \emph{n}-mers depends non-trivially on both the ion valency and size. As the valency is increased, the ion-water bonds are strengthened. However, the ion size contributes to determine at which separation the hydration shells are forced to reduce in size, with smaller ions being able to carry their bulk-like hydration structure to smaller separations. In any case, both the ion size and valency control the layering in a complex way because higher valency ions strongly prefer to both remain bound to water and to repel other ions, while larger ions are restricted from separating into two layers at lower separations due to steric repulsion. As we have seen, the energetic preference to maintain a larger hydration shell can, at certain separations, outweigh the electrostatic drive for the ions to arrange into two layers.

These competing effects are clear for Al\(^{3+}\) ions, which are the smallest and have the greatest charge. Both of these two attributes allow the 6-mers to persist even at a separation of 7\si{\angstrom}, thanks to adaptations in shape and layering. Because the other two ions are larger and have comparatively weaker electrostatic interactions, they are unable to maintain their bulk-like hydration at such a separation and therefore change to lower water coordination: this leads to a relatively stronger net attraction for lower valency ions. A similar net attraction with monovalent ions under high confinement has been recently demonstrated in recent neutron and x-ray experiments \cite{mukhina2019attractive} and discussed in fully atomistic simulations \cite{Shen2021,brochard2021swelling}.

The effect of ion hydration noted here is also consistent with the presence of repulsive oscillatory hydration forces detected in experiments on the interaction between mica surfaces in ionic solutions\cite{israelachvili2015intermolecular,israelachvili1983molecular, pashley1984molecular}, which was attributed simply to the layering of water molecules. However, as postulated by more recent experiments \cite{espinosa2012hydrated,zachariah2016stepwise,zachariah2017ion}, the shifts in these forces are rooted in layering transitions of hydrated ions, or film-thickness transitions, which are ion-specific and depend on the size and strength of the hydration shells. At low surface charge densities, we find a similar interplay of ion-water coordination and ion layering is behind the oscillations in the surface-surface force.
Our semi-atomistic approach demonstrates that non-monotonic, ion-specific effects in the nanoscale forces between charged surfaces can emerge from just the changes in ion-ion and ion-water correlations due to the size and valency of ions, which may be at the origin of Hofmeister effects in clays, and potentially in wider contexts.

\begin{figure*}
\centering
\includegraphics[width=6.5in]{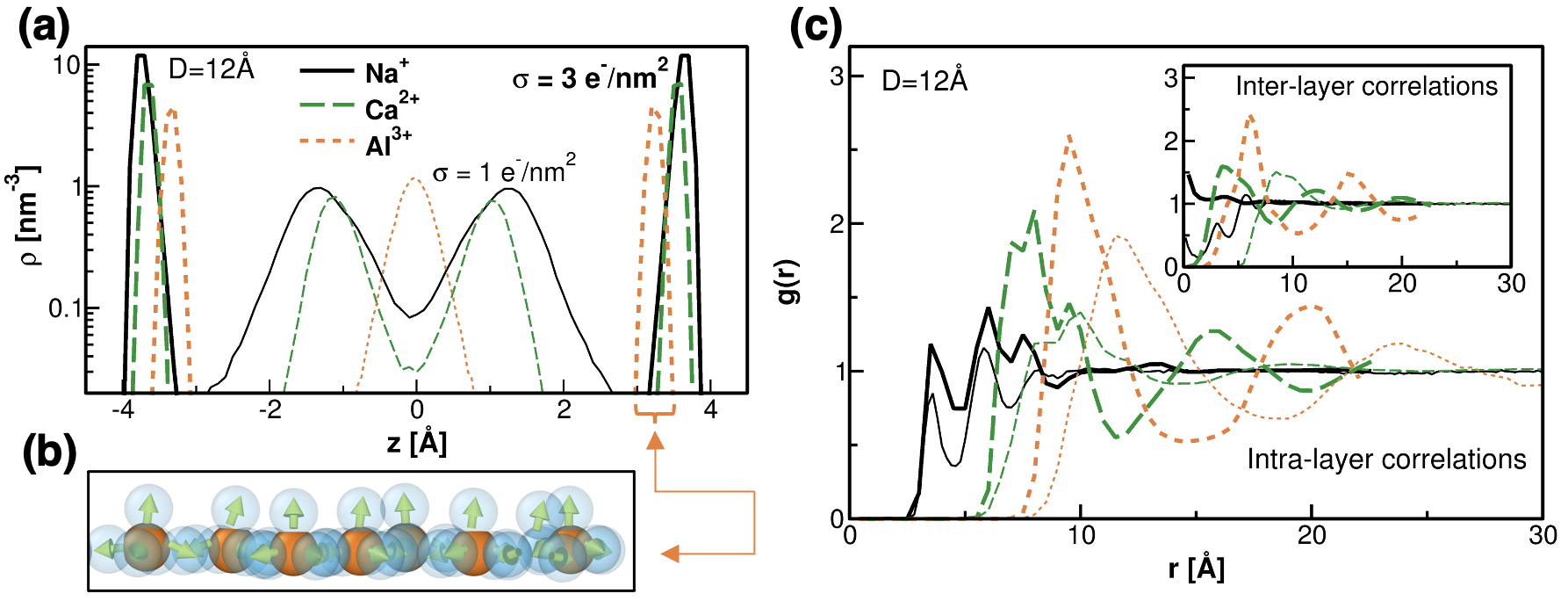}
\caption{When the surface charge density is increased to \(\sigma = 3e^{-}/nm^{2}\), all of the ions localize into two layers adjacent to the surfaces at every value of D, as shown by the density profiles at D=12\si{\angstrom} (left). The accompanying simulation snapshot of aluminum ions in one of these layers illustrates that this new arrangement forces the hydration shells to be hemispherical in shape. On the right, at larger sigma the ions within each layer are pushed closer to one another in the xy plane, resulting in intensified intra-layer and inter-layer correlations, both of which grow with the valency. }
\label{figure5}
\end{figure*}

\subsection*{Increasing the Surface Charge Density}

Increasing the surface charge density from \(\sigma = 1e^{-}/nm^{2}\) to \(\sigma = 3e^{-}/nm^{2}\) allows us to investigate how the ion and water structures which control the ion-specific forces may be affected. To maintain the system's electrostatic neutrality while the total number of ions is kept fixed, the area of the plates at \(\sigma = 3e^{-}/nm^{2}\) is smaller than at \(\sigma = 1e^{-}/nm^{2}\). All ions react to the stronger electrostatic forces in this condition by strongly localizing into two layers pressed against the charged surfaces, as illustrated by the density profiles in Figure \ref{figure5}a. Studies building Hofmeister series for clays have highlighted the competition between the ion-surface adsorption energy and the energy of their bulk hydration shell \cite{underwood2016ion}.  The closest distances observed in these density profiles are consistent with that. As a result, for this surface charge density, the ions are so close to the surfaces, at all separations, that they cannot have full hydration shells (see Figure \ref{figure5}b). This effect of the higher $\sigma$ agrees with simulations of silica nanopores, which have also revealed that increasing the surface charge density pushes the ions closer to the surfaces and reduces the hydration coordination number of the ions \cite{collin2018molecular}. The bounded water molecules are also generally less dynamic at higher surface charge density (Figure S5 in the supplementary information shows Al\(^{3+}\) as an example). Importantly, these hemispherical hydration shells have less energy than the full hydration shells found at $\sigma=1e^{-}/nm^{2}$, and are thus more susceptible to destabilize when the confinement increases.  

\begin{figure*}
\centering
\includegraphics[width=6.5in]{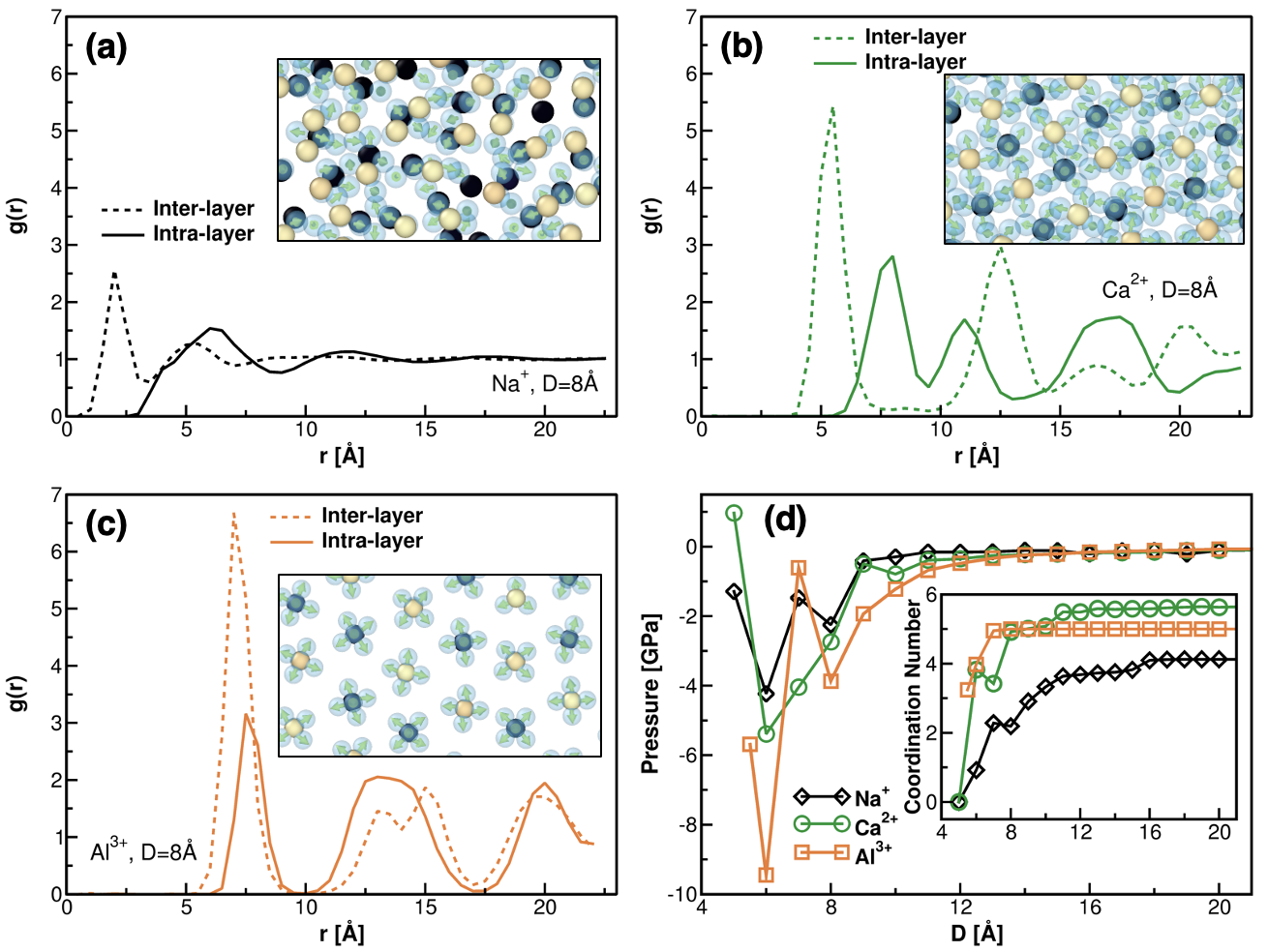}
\caption{Inter-layer and intra-layer correlations at $\sigma=3e^{-}/nm^{2}$ and D=8\si{\angstrom} for sodium (a), calcium (b) and aluminum (c). Reducing the pore size to D=8\si{\angstrom} leads to higher peaks in the $g(r)$ when compared to Figure \ref{figure5}c; the inter-layer correlations especially increase. This is also reflected and visualized in the corresponding simulation snapshots. The growth of inter-layer correlations gives rise to a greater attraction between the surfaces (d). Nevertheless, the stability of the hemispherical hydration shells still has an effect; the few spikes in the pressure are rooted in changes in the coordination number, or size of the \emph{n}-mers (inset). }
\label{figure6}
\end{figure*}

Increasing the surface charge density also modifies the behavior of the ions within the two layers in the $\hat{x}$ and $\hat{y}$ directions; the pair correlation function in Figure \ref{figure5}c illustrates that the ions are now closer to one another in each layer. Additionally, the intra-layer and inter-layer correlations (when there are two layers) are stronger at higher $\sigma$ and, once again, grow with the valency. Furthermore, ion positions become even more correlated as the pore size is reduced from D=12\si{\angstrom}, especially at higher $\sigma$ and higher valency. In particular, at D=8\si{\angstrom}, the ions become dramatically more correlated across layers (Figure \ref{figure6}). Visually, for aluminum, the structure of the ions becomes lattice-like at this value of D, while calcium ions, being more numerous and packed more tightly, form a similar but staggered lattice \cite{Samaj2012,Samaj2018}---i.e. with neighboring ions in opposite layers (Figure \ref{figure6}b). For sodium ions, the case with the weakest electrostatic coupling, the inter-layer correlations are indeed intensified at small D, but clearly not to the extent, in range or magnitude, of the other ion species. 

The stronger ion correlation forces at $\sigma=3e^{-}/nm^{2}$, coupled with the fact that less energy is needed to dehydrate the hemispherical \emph{n}-mers, translate into a stronger attractive force between the two charged surfaces (Figure \ref{figure6}d). The net force at higher $\sigma$ also features less oscillations, as there are fewer changes in \emph{n}, and no layer transitions. Nevertheless, as highlighted by the inset of Figure \ref{figure6}d, the \emph{n}-mer stability still plays a role and explains the spike in pressure for Al\(^{3+}\) at D=7\si{\angstrom}. The energy required here to remove the 5th nonplanar bound water molecule, visualized in Figure \ref{figure5}b, is the source of this increase, similar to the large spike seen in the pressure at $\sigma=1e^{-}/nm^{2}$. Water polarization (or ionization, which is likely with Al\(^{3+}\)) would probably affect the exact energetic-entropic balance for \emph{n}-mers compared to the relatively simple water model used in this study---as also would molecular details of the surface such as charge heterogeneity or roughness. While all these additional factors would affect some of the reported details on cohesion strength, \emph{n}-mer populations, etc., it is remarkable that a clear inter-dependence emerges even without those effects, providing a possible explanation for the fundamental mechanisms underlying ion-specificity in a wide range of clays and other systems.

\section*{Conclusions}

Using a semi-atomistic approach, we have demonstrated the importance of using explicit water molecules to describe ion-specific surface interactions for ionic solutions confined between charged surfaces, relevant to clays. The introduction of explicit water sheds new light into how the net pressure between two charged clay surfaces depends non-trivially on the counterion type and the surface charge density, especially at strong confinement. Under extreme confinement, the water provides minimal electrostatic screening, inducing a net attraction for all ions (even Na\(^{+}\)) at most pore widths. However, at surface charge densities typical of smectite clay surfaces, the ion-specific pressure also fluctuates strongly as the ion-specific size, and shape in the case of Al\(^{3+}\), of the hydration shells changes with the pore width. These oscillations can be significant in magnitude and induce a net repulsive pressure, especially when the hydration shells are highly stable (such as with Ca\(^{2+}\) and especially Al\(^{3+}\)) and are forced to change with increasing the confinement. At higher surface charge density, the ion-water structuring becomes even more correlated and results in frozen-like configurations, leading to a highly attractive net pressure that increases with the ion valency. Nevertheless, even at higher surface charge density, changes in the ion hydration shells have a significant impact on the surface-surface interactions. Overall, ion-surface interactions, surface-surface interactions, and ion specificity effects in clays are ultimately controlled by the different structures that the ions form with the water in confinement. 
The interaction strengths and ion specificity investigated here have dramatic consequences at much larger lengthscales in real clay materials: it determines interlayer distances between nanoscale clay particles, depending on water content, salinity and ionic species in specific contexts. Moreover, the shape of the net interactions and the presence of competing attraction and repulsion at the nanoscale also has dramatic implications for larger lengthscales, because it determines the anisotropic growth of aggregates into fibrils, lamellae, and layered mesophases that then self-assemble into gels and large scale porous structures \cite{deCandia2006columnar,ioannidou2016crucial,bourg2017clay,Goyal2020,Shen2021}. Understanding and predicting these features, therefore, is the first step to obtain the missing link from the nanoscale to the mesoscale aggregation kinetics and morphological variability of clay soils and clayey materials.  Eventually, the nanoscale forces and the resulting mesoscale aggregates determine the development of larger pores and of local stresses in the final larger scale matrix, and the coexistence of compressive or tensile stresses, which have consequences for the long term evolution and the interactions with the environment of clay sediments \cite{deshpande2021perpetual,bourg2017clay,hochella2019natural,brochard2017nanoscale,brochard2021swelling}.

To conclude, our results clarify the mechanism by which ionic composition controls the properties of hydrated clays and open the path to understand the complex larger scale behavior of clay-based materials. Moreover, the same mechanism may help explain Hofmeister effects in a wider array of systems, from macromolecular solutions to proteins, membranes, and colloids.

\section{Acknowledgements}
The authors acknowledge the NIST PREP Gaithersburg Program (70NANB18H151), the Georgetown Undergraduate Research Opportunity Program and the NASA DC Space Grant Consortium for support. 

\section{Supporting Information}
Mean-square displacement of ions in xy plane and perpendicular z direction with PM and EW; mean-square displacement of bound and free water in xy plane and z direction; self-intermediate scattering function of ions, bound water and free water 

\bibliography{References}
\clearpage

\end{document}


\begin{figure*}
\centering
\includegraphics[width=5in]{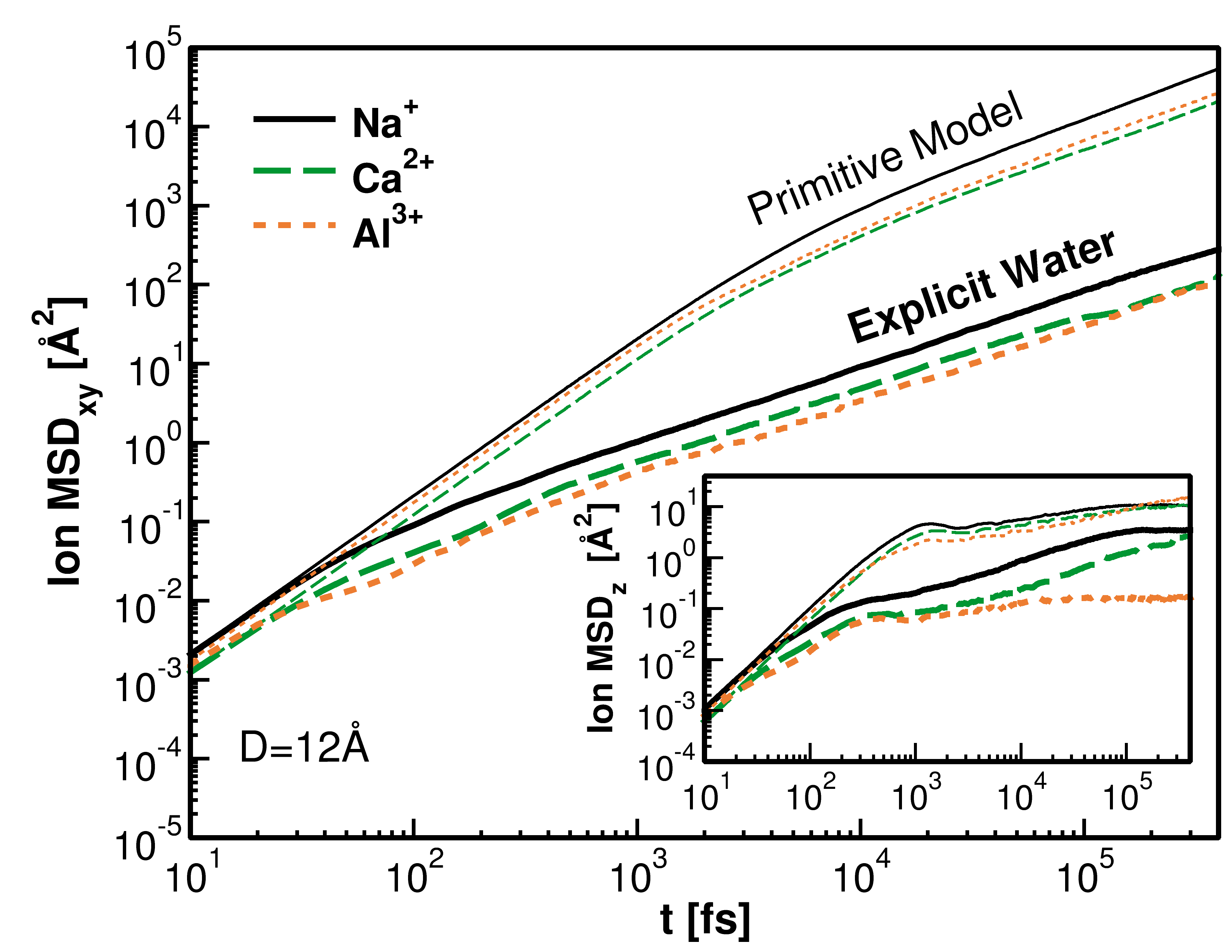}
\caption{The mean-square displacement of the ions in the plane parallel to the surfaces (xy) and in the perpendicular z direction (inset), for $\sigma=1e^{-}/nm^{2}$ and D=12\si{\angstrom} (the same system conditions as the ion profiles of figure 2). Utilizing explicit water (EW) slows down the ions in all directions, with Na\(^{+}\) being the most dynamic for both models. The changes in layering that result from including EW, as described in the main text, also affect ion mobility; Al\(^{3+}\) ions in explicit water, being squeezed into a single layer at D=12\si{\angstrom}, are considerably more localized along the z-direction. The other two ions, and all of the ions in the primitive model (PM), are able to switch between the two layers. }
\label{figureS1}
\end{figure*}

\begin{figure}
\centering
\includegraphics[width=5in]{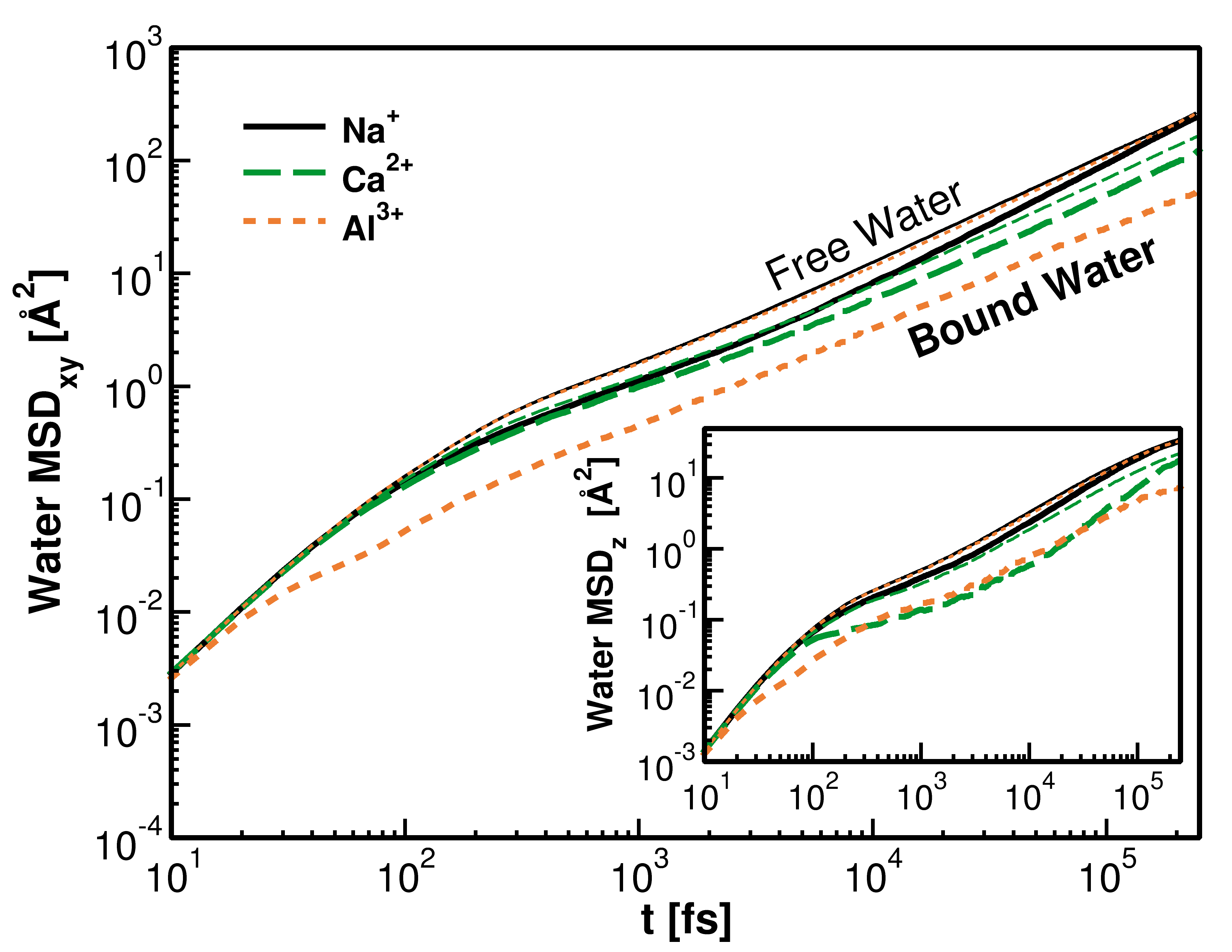}
\caption{The mean-squared displacement in the plane parallel to the surfaces and along the perpendicular z direction (inset) for the bound (thick) and free (thin) water, at D=20\si{\angstrom} and $\sigma=1e^{-}/nm^{2}$. Water bound to sodium displays dynamics not markedly different from the unbound water, due to the weaker ion-water bonds depicted in figure 3a of the main text. As the valency is increased, the mobility of the bound water becomes increasingly coupled to the slower ions. }
\label{figureS2}
\end{figure}

\begin{figure*}
\centering
\includegraphics[width=6.6in]{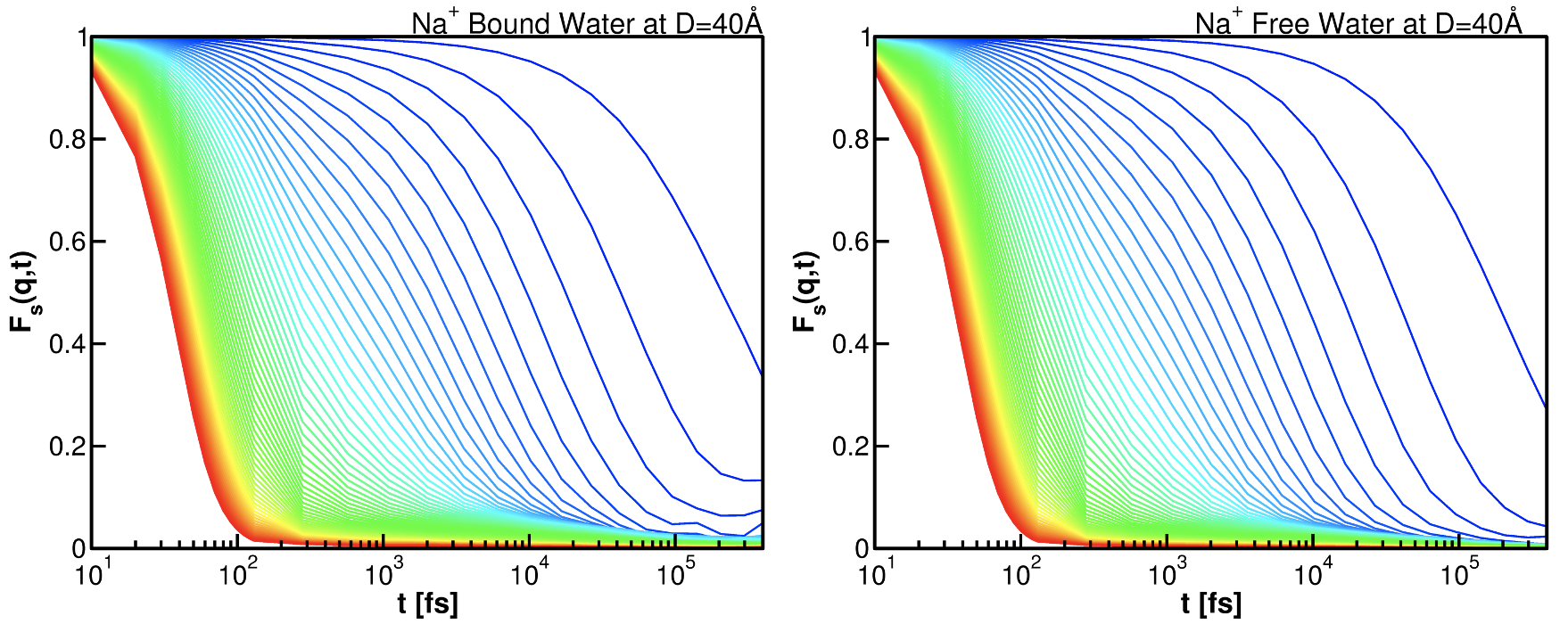}
\caption{The self-intermediate scattering function computed for water initially bound to Na\({+}\) ions and free water, at D=40\si{\angstrom} and $\sigma=1e^{-}/nm^{2}$. In this and subsequent figures, the value of \(q_{z}\) ranges from  \(q_{z} = 2 \pi / D\) (blue), to \(q_{z} = 10 \si{\angstrom}^{-1}\) (red). The dynamics of the water physically bound to the sodium ions does not differ much from the free water mobility. This small-factor difference is consistent with quasielastic neutron scattering (QENS) experiments on hydrodynamics in smectite clays, which have similarly charged surfaces and monovalent interlayer ions \cite{swenson2000quasielastic, michot2012anisotropic, marry2013anisotropy}. }
\label{figureS3}
\end{figure*}

\begin{figure*}
\centering
\includegraphics[width=6.6in]{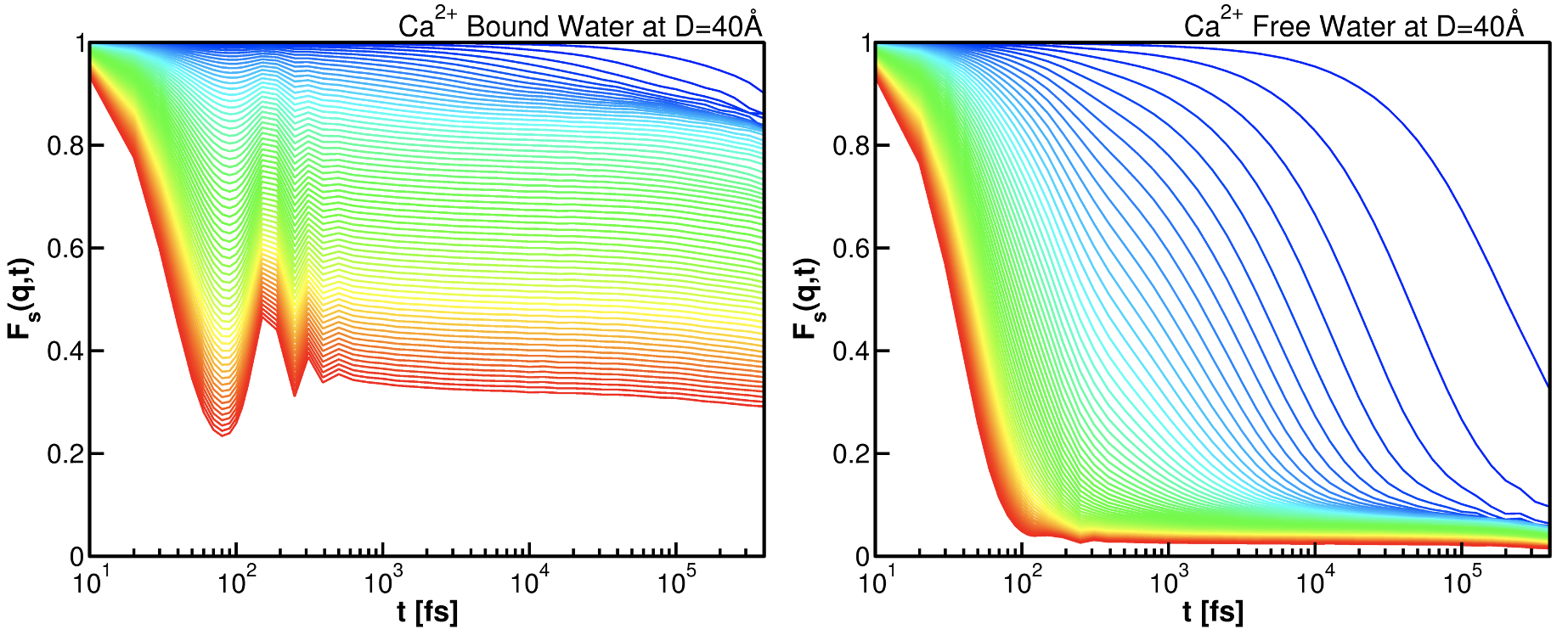}
\caption{For higher electrostatic coupling, the dynamical differences between bound and free water becomes more pronounced, concurring with the stronger ion-water bond correlations found at higher valency (see main text). With calcium counterions and $\sigma=3e^{-}/nm^{2}$, conditions typical of hydrated cement, the motion of the water that hydrates the ions is clearly more constrained than the bulk water, as illustrated by the plateaus in the SISF for the bound water. This contrast is also in agreement with the distinction between slower ``physically bound" water and free water, made in QENS observations of cement \cite{bordallo2006water,thomas2001state}. This distinction is also seen in Inelastic Neutron Scattering studies of smectite minerals with interlayer Mg\(^{2+}\), but not with ions with a lower hydration enthalpy, which is in agreement with figure \ref{figureS3} \cite{cygan2015inelastic}. }
\label{figureS4}
\end{figure*}

\begin{figure*}
\centering
\includegraphics[width=6.6in]{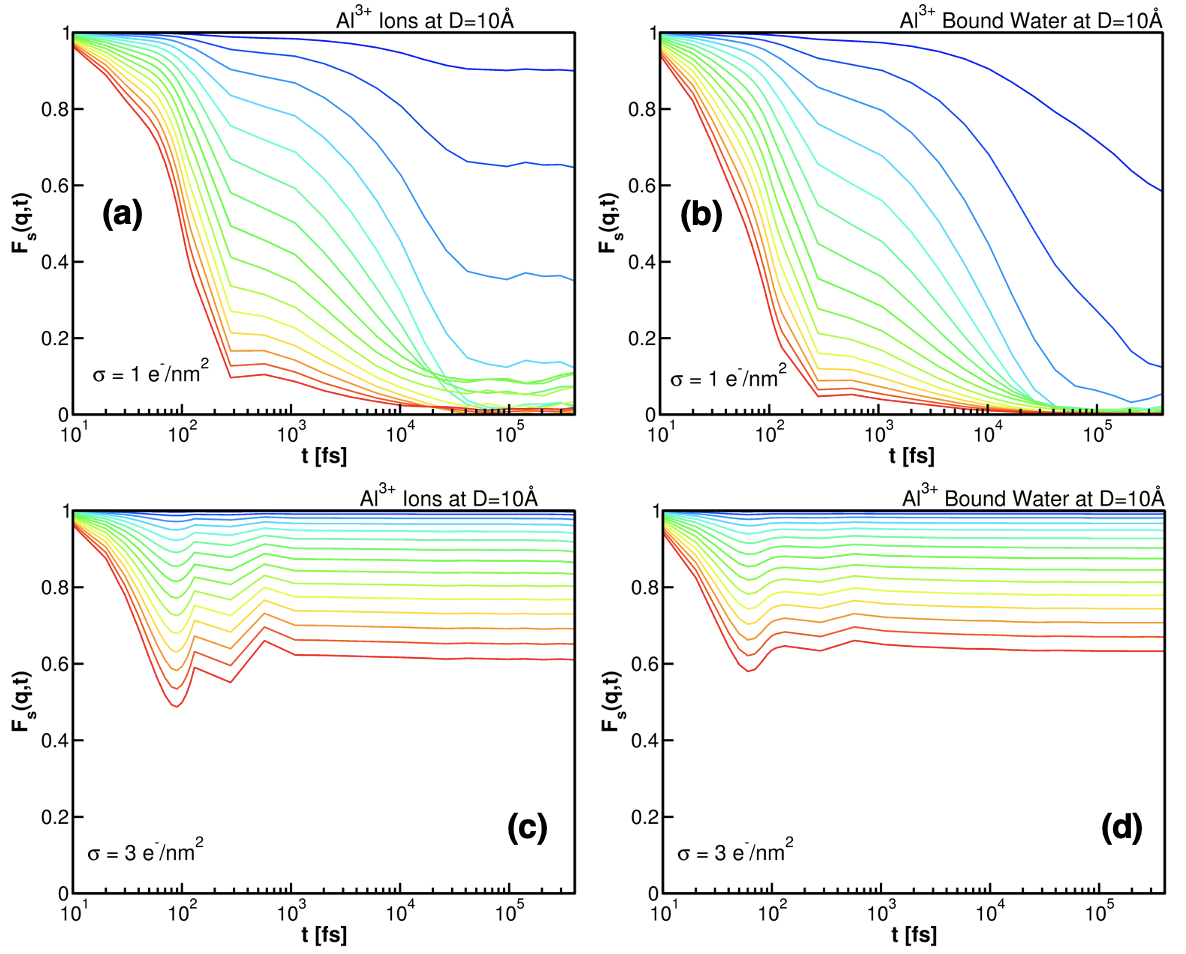}
\caption{SISF at D=10\si{\angstrom} and $\sigma=1e^{-}/nm^{2}$ for Al\(^{3+}\) ions (\textbf{a}) and bound water (\textbf{b}), with the same respective quantities computed at $\sigma=3e^{-}/nm^{2}$ (\textbf{c, d}). Even though Al\(^{3+}\) \emph{n}-mers remain stable for the whole simulation regardless of the value of D or $\sigma$, the motion within the hydration shells does vary with both. The ions and bound water at $\sigma=3e^{-}/nm^{2}$ are strongly localized, while the bound water at $\sigma=1e^{-}/nm^{2}$ is more dynamic and free to rotate about the ions, which eventually localize in the z-direction. }
\label{figureS5}
\end{figure*}


\clearpage
\bibliography{referencesSupp}